\documentclass[11pt]{article}
\input{epsf}

\def\be{\begin{equation}}
\def\ee{\end{equation}}
\def\ba{\begin{eqnarray}}
\def\ea{\end{eqnarray}}

\def\12{{1\over 2}}

\def\msun{M_\odot}

\def\ltsima{$\; \buildrel < \over \sim \;$}
\def\simlt{\lower.5ex\hbox{\ltsima}}
\def\gtsima{$\; \buildrel > \over \sim \;$}
\def\simgt{\lower.5ex\hbox{\gtsima}}

\begin{document}
% \selectlanguage{english}

\title{\bf Contribution of HD molecules in cooling of the primordial gas}
\author{E.~O.~Vasiliev$^{1,2}$, Yu.~A.~Shchekinov$^2$\thanks{yus@phys.rsu.ru} \\
\it $^1$Institute of Physics,
\it $^2$Department of Physics, Rostov State University,\\
\it Rostov-on-Don, 344090, Russia}

\date{}

\maketitle

\begin{abstract}
We study the effects of HD molecules on thermochemical evolution of the
primordial gas
behind shock waves, possibly arised in the process of galaxy formation.
We find the critical shock velocity when deuterium transforms efficiently
into HD molecules which then dominate gas cooling. Above this
velocity the shocked gas is able to cool down to the temperature of the
cosmic microwave background. Under these conditions the corresponding Jeans
mass depends only on redshift and initial density of baryons
$M_J \propto \delta_c^{-0.5} (1+z)^{0.5}$. At $z\simgt 45$ HD molecules heat
shocked gas, and at larger redshift their contribution to thermal evolution
becomes negligible.

\end{abstract}

%\newpage

%----------------------- Section 1 -------------------------------

\section{Introduction}

\noindent
Formation of stars is intimately connected with the ability of gas to cool.
In the metal-free primordial medium the radiative cooling is
mainly provided by molecular hydrogen and its isotope analogue HD. In the expanding
universe H$_2$ and HD form after the epoch of re\-com\-bina\-tion
[1-5]. Due to a non-zero dipole moment and lower
excitation energy HD molecules can efficiently cool gas at temeperature
$T\le 200$~K, where the rate of H$_2$ cooling decreases sharply.
If HD abundance is small, cooling of primordial gas stops practically at $T\leq 200$ K.

Thus, thermal evolution of the primordial gas at low temperature,
and as a consequence
characteristics of the first stars in the universe critically depend on
the abundance of HD molecules. At present, an analysis of conditions when HD
can be thermodynamically important in the primordial gas is absent.
Conclusions about the role of HD molecules in a
prestellar universe are contradictory. In particular, it is shown in \cite{lepp84}
that HD cooling never dominates in a collapsing spherical cloud.
At the same time the authors [6-8] point out that HD cooling
can be important in primordial clouds, although their calculations
are restricted only by initial stages of the collapse.
It is obvious though, that the abundance of HD crucially
depends on thermal evolution of gas in the temperature range $T>500$ K.
The apparent contradiction about the role of HD
is connected with differences in the initial conditions used in \cite{lepp84}
from one side and in [6-8] from the other. In addition,
the H$_2$ cooling function adopted in \cite{lepp84} is overestimated. Therefore, in
order to make firm conclusions about the role of HD molecules in a prestellar
universe it is necessary to investigate the efficiency of HD formation
in a wider range of initial conditions.

It is known that formation of H$_2$ and HD molecules is largely
boosted behind shock waves [10-13].
It is connected mostly with the fact that temperature and fractional ionization
behind the shock waves increase, and as a consequence
the rates of molecular reactions are enhanced. In a postshock gas
cooled down to temperature $\sim 10^4$ the fraction of electrons
remains sufficiently high, $x\simgt 0.001$, which favours
rapid formation of H$_2$ molecules in the catalystic reactions
with H$^-$ ions. At such conditions H$_2$ fraction can reach $\sim 10^{-2}$.
Further cooling is mainly provided by H$_2$ molecules efficient
in the temperature range $200-7000$~K. When lower temperatures
are reached, $T\le 200$~K, deuterium begins to convert into HD molecules due to
chemical fractionation \cite{solomon,varsh}. The
contribution of HD cooling in energy losses increases when temperature decreases,
and if it becomes dominant the gas temperature can fall down
to several tens of degrees. One can expect that at least in a restricted
range of shock parameters formation of HD molecules is as efficient as to
provide such a predominance. We aim to study this possibility.

In Section 2 we describe a thermochemical model of a shocked gas;
the results are presented in Section 3; in Section 4 discussion and
in Section 5 summary are given. Throughout the paper we assume a
$\Lambda$CDM cosmology with the parameters
$(\Omega_0, \Omega_{\Lambda},\Omega_m,\Omega_b,h)=(1.0,0.71,0.29,0.047,0.72)$
and deuterium abundance $2.62\times 10^{-5}$ \cite{wmap}.

%----------------------- Section 2 -------------------------------

\section{Molecular kinetics behind shock waves}

\noindent
In the absence of thermal conductivity and diffusion the thermochemical
evolution of gas behind a shock wave can be described by
a system of ordinary differential equations for a Lagrangian fluid element,
which include equations of chemical kinetics
\be
\label{spe}
\dot x_i = F(x_i,T,n) - D(x_i,T,n),
\ee
and energy equation
\be
\label{temp}
\dot T={2\over 3}\sum\limits_i[\Gamma_i(x_i,T,n) - \Lambda_i(x_i,T,n)] + {2\over 3}{T\over n}\dot n,
\ee
where  $x_i$ is the fraction of $i$-th species, $F_i({\bf x},T,n)$,
$D_i({\bf x},T,n)$ are the corresponding formation and destruction rates,
$\Gamma_i({\bf x},T,n)$, $\Lambda_i({\bf x},T,n)$ are the
heating and cooling rates. Chemical kinetics of primordial gas
include the following main species: H, H$^+$, H$^-$, He, He$^+$, He$^{++}$,
H$_2$, H$_2^+$, D, D$^+$, D$^-$, HD, HD$^+$, $e$. The rates for collisional and
radiative processes are taken from \cite{gp,shapiro}, and for D$^-$ ion
from \cite{sld98}. The energy equation accounts cooling processes connected with
H, He, He$^+$, He$^{++}$, such as collisional excitation and
ionization, bremsstrahlung radiation, recombination, dielectronic
recombination, molecular cooling by H$_2$ and HD, and Compton cooling.
In the absence of external ionizing radiation
the abundances of chemical species and radiative
cooling are determined by collisions. Thus, the right-hand side
of (1) and first term of the right-hand side of (2) are
proportional to the gas density, and for convenience
we can introduce the fluence $\eta$ through
\be
\label{fluence}
d\eta = n dt,
\ee
where $t$ is time, $n$ is number density; below the results are presented
as functions of the fluence.

For the H$_2$ cooling functions we adopted the expressions given in \cite{hm},
the HD cooling is taken from \cite{flower}, the other rates are from \cite{cen92}.
In addition, the effects of interaction between molecules and the CMB photons are
accounted \cite{puy93,varsh,gp00} which imply that when gas temperature
is close to the CMB temperature, H$_2$ and HD molecules are populated
by the CMB photons, and then collisionally transfer the energy to kinetic temperature,
thus heating the gas. Therefore, gas cannot cool lower than the CMB temperature.
We assume that the gas temperature behind the front jumps and reach the value
\be
\label{tshock}
T_0 = \alpha^2 {m_p v_c^2 \over k} \simeq 1.2\times 10^2~\alpha^2
      \left({v_c \over 1 {\rm km}~{\rm c^{-1}}}\right)^2,
\ee
where $\alpha^2 = 3/16$ is for a shock propagating in a static gas,
and $\alpha^2 = 1/3$ is for a shock wave formed in a head-on collision of
two flows (clouds) with equal velocities  $v_c$ \cite{smith}.
Neglecting thermal conduction the evolution of each Lagrangian volume of gas behind
the shock is isobaric \cite{coll,shapiro,annorman},
so that the density is described by
\be
\label{iso}
n={p\over \mu kT},
\ee
where $\mu$ is the average molecular weight.

%----------------------- Section 3 -------------------------------

\section{Formation of HD behind shock waves}

\noindent
In the contemporary scenarious of structure formation in the universe
the first protogalaxies form at the epoch $z=10-30$. For specifity,
we consider thermochemical evolution of baryons behind a
shock at the redshift $z=20$. Formation of dark haloes
(future protogalaxies) and their subsequent virialization are accompanied
by shock waves in the baryon component. The duration
of this process is close to the Hubble time $t_H(z)$ \cite{zeldnov},
so we restrict computations by $t_H(z)$, which means the ending redshift $z_e \simeq 12$
for the initial $z_i=20$.
One can expect that collisions of baryonic flows during the virialization
of dark haloes result in considerable density variations. In order to study possible
influence of such density variations on thermochemical evolution
we conduct the calculations for a wide range of density in colliding flows:
from the background to the virial value. As a characteristic value we adopt
the virial density, $18\pi^2n_b(1+z)^3$ (see e.g. \cite{t97}), where $n_b$ is
the background baryon density today, and consider the dependence of the
thermochemical evolution on the initial gas density.
Gas before the shock is assumed to be cold compared to the gas just
after the front, so we consider strong shock waves.

HD molecules form efficiently at low teperature in the presence of
sufficient fraction of molecular hydrogen through the reaction
\be
\label{hdform}
{\rm D^+ +H_2 \longrightarrow HD+ H^+}.
\ee
For this reason we briefly discuss formation and destruction of H$_2$ molecules
behind shock fronts \cite{coll,shapiro,kang,ohhaiman}.

In the primordial gas H$_2$ forms in interactions
of neutral hydrogen with H$^-$ and H$_2^+$ ions efficiently born
at high temperature. As shock waves significantly increase
temperature it enhances formation of H$_2$ and increases
its abundunce \cite{shapiro}. It is known that HI
cooling becomes inefficient at temperature $\simlt 10^4$~K,
and in the primordial gas only the molecular hydrogen can provide further cooling
to lower temperature.
One can estimate a minimum fraction of H$_2$ needed for effective
cooling: the cooling time must be shorter then the Hubble time
which condition fulfills when H$_2$ fraction becomes greater than the critical value
$x_{\rm H_2} = 5\times 10^{-4}$ \cite{t97}; an increase of H$_2$
abundance shortens the cooling time. In a shocked gas
at temperature $\simgt 8\times 10^3$~K
collisions increase fractional ionization,
and this further enhances formation of H$^-$, H$_2^+$ and H$_2$. Figures 1-3
show the evolutionary paths of the thermochemical state of a gas element
behind the shock: the evolution begins at high temperature and
follows a monotonous cooling accompanied by grownig H$_2$
and HD fractions. It is seen that at high collisional velocities
the electron fraction significantly increases at initial stages,
which stimulates formation of H$^-$ ions and H$_2$ molecules
and a strong decrease of temperature. The H$_2$ abundance grows rapidly
in the temperature range $T\sim 10^3-10^4$~K, while at lower temperatures
formation of H$_2$ practically exhausts \cite{ohhaiman}. Already formed H$_2$
molecules provide further cooling down to $T\sim 200-400$~K depending
on the exact value of their abundance. If gas temperature
falls down to $T\simlt 150$ K deuterium rapidly converts to HD molecules
due to chemical fractionation (Fig.4.).

Fig. 5 presents the dependence of electron, H$_2$ and HD fractions on
the fluence $\eta=\int ndt$ for several values of the shock velocity.
Note that for the shock waves with $v_s \geq 3.5\alpha^{-1}$~km~s$^{-1}$
the final abundance of H$_2$ is greater than the limit $5\times 10^{-4}$,
and as a cosequence the gas behind the shock can lose significant
fraction of its thermal energy in one Hubble time. Since larger
shock velocity corresponds to higher temperature, the maximum density
behind the shock increases with the velocity $\rho\propto v_s^2$
(eqs. \ref{tshock} and \ref{iso}). In Fig.5 this corresponds to
larger fluence. It is obvious, that the characteristic time of the thermal
evolution decreases as $\propto v_s^{-2}$.
For $v_s \simeq 4.6\alpha^{-1}$~km~s$^{-1}$ H$_2$ fraction equals
$\sim 7\times 10^{-4}$ and becomes sufficient for cooling down to
$T\simeq 130$~K; HD fraction at these conditions is $4\times 10^{-7}$.
Further increase of the shock velocity results in an increase of
molecular fractions and a decrease of temperature on a shorter time,
with the dominant cooling provided by HD molecules (Fig.6). Equating
the cooling rates from H$_2$ and HD one can estimate the critical
temperature below which the contribution of HD into thermodynamics
becomes dominant. For the adopted cooling functions it occurs
at $T_{\rm cr}\simeq 130$~K. For the shock wave velocity
$\sim 4.6\alpha^{-1}$~km~s$^{-1}$ gas cools down to this limit, and for
larger velocities temperature falls below $T_{\rm cr}$ where the HD cooling
dominates.

In the velocity range $4.6\alpha^{-1} \simlt v_s \simlt 8.7\alpha^{-1}$~km~s$^{-1}$
the initial temperature behind the shock is insufficient for collisional ionization,
and in the subsequent evolution the electron fraction can only decrease
(Figs.1 and 5). At such conditions H$_2$ and HD abundances
increase with velocity only because the reaction rates grow with temperature.
As seen in Fig.2 for shock velocities in this range
the maximum fraction of H$_2$ is approximately equal to
$x_{\rm H_2} \simeq 8\times 10^{-4}$. This is enough for an efficient formation
of HD molecules and successful cooling down to $T\le 130$~K.
At $v \simgt 7\alpha^{-1}$~km~s$^{-1}$ the HD fraction becomes sufficient
for cooling down to the CMB temperature $T_{\rm CMB} = 2.7(1+z)$: due to the strong
emission in rotational lines of HD gas temperature falls to several
tens, $\sim 30$~K, approaching $T_{\rm CMB}(z)$
at a given redshift (Fig.5). This is because HD molecules provide
an efficient exchange of energy between the CMB and baryons through
absorption of CMB photons and subsequent collisional de-exitation
\cite{puy93,varsh,gp00}. It is obvios that similar picture
is valid at all redshift, and the final temperature of cold baryons
is $T_{\rm min}\simeq T_{\rm CMB}$.

For velocities $v_s \simgt 9.2\alpha^{-1}$~km~s$^{-1}$
gas temperature behind the shock becomes greater than $10^4$~K,
which results in an increase of fractional ionization immediately
after the gas element
crosses the shock front: for $v_s = 10.4\alpha^{-1}$~km~s$^{-1}$ $x_e$
increases by factor of 2, and for $v_s \simgt 11.6\alpha^{-1}$~km~s$^{-1}$
more than an order (Fig.1 and 5). At such conditions chemical kinetics
changes qualitatively -- the enhancement of H$_2$ formation
is caused in this case by the two factors: increasing reaction rates and
a higher ionization fraction, resulting in more frequent catalystic processes
H$+e\to$H$^{-1}$, H$^{-1}+$H$\to$H$_2+e$. The corresponding
evolution of $x_e(t)$, $x_{{\rm H}_2}(t)$ and $x_{\rm HD}(t)$ looks
qualitively different as seen in Fig.5: while for $v_s<9.2\alpha^{-1}$~km~s$^{-1}$
an increase of velocity by 1~km~s$^{-1}$ produces insignificant changes
in $x_{{\rm H}_2}(t)$ and $x_{\rm HD}(t)$, for greater
velocities such an increase results in a considerable
(half order of magnutude) increase of $x_{{\rm H}_2}(t)$ and $x_{\rm HD}(t)$.
Thus, $x_{{\rm H}_2}(t)$ and $x_{\rm HD}(t)$
fractions at equal temepratures are higher for larger velocities (Figs. 2 and 3).
For lower velocities the evolutionary paths
$x_{{\rm H}_2}(T)$ and $x_{\rm HD}(T)$ for different $v_s$ practically
coincide at temperature $T\leq 10^3$ K. One should stress, that
in the considered range of velocities H$_2$ molecules form primarely
through H$^-$ ions, the contribution from H$_2^+$ ions is as a rule negligible.

Thus, for velocities $v \geq 4.6\alpha^{-1}$~km~s$^{-1}$ HD molecules
behind shock waves provide lower temperatures than
H$_2$ can do. It is worth noting that at $z \sim 45$
the CMB temperature is higher than the critical value $T_{\rm cr}$ at which
HD cooling dominates. Therefore, under these
conditions HD molecules can only heat gas, and at larger redshifts become
unimportant.

Everywhere above we adopted gas density in colliding flows equal to the virial value
at the corresponding redshift. However, one can assume that
in the process of merging of haloes a fraction of baryonic mass can be lost. In the intergalactic
medium such ''separated'' baryonic flows can greatly expand, and their final density
depends on collision velocity $v_c$, masses of the merging haloes, details of
separation and so on. Subsequently such baryonic clumps can
collide with gaseous components of other haloes or with each other. In these
conditions the thermal evolution differs from that of denser baryonic flows. Let us
consider how the thermochemical evolution depends on the density. Fig.7
shows the temperature and the HD fraction versus the density for several shock velocities at two
redshifts. It is clearly seen that at $z = 20$ and in the low velocity range $v\leq 5.8\alpha^{-1}$~km~s$^{-1}$ only for densities close to the
virial value gas temperature drops below the critical value where
contribution from HD dominates. However, for higher velocities HD
cooling remains efficient even for
densities of one order of magnitude lower than the virial value. At $v\simeq 5.8\alpha^{-1}$~km~s$^{-1}$
and the density close to the virial value only $\simeq 0.25$ of deuterium converts in HD,
however it becomes sufficient to cool the gas down to the CMB temperature.
For higher velocities
this can occur for several times lower densities than the virial value. Collisions
with higher velocities $v \geq 11.6\alpha^{-1}$~km~s$^{-1}$ change H$_2$ kinetics:
formation of H$_2$ becomes more efficient due to significant increase of
fractional ionization behind the front, and behaves similar to the collisions
with virial density at $v_s \simgt 10.4\alpha^{-1}$~km~s$^{-1}$ shown in Figs. 1 and 5.
These features are seen in Fig.7. In general, one can conclude that HD molecules can
also play a significant role in cooling of baryonic flows of low density.

\section{Discussion: formation of protostellar fragments}

\noindent
Birth of stars is always accompanied with shock waves. This is
unconditionally true for the first stars in the universe. The origin of shock waves
can be connected both with the formation of the first protogalaxies
in merging flows, and with supernovae explosions in already formed galaxies.
Formation of the first protogalaxies implies separation of highly overcritical
density perturbations from the Hubble expansion and a predominantly
one-dimensional compression \cite{lin,zeld}. This is a source of shock waves
in baryonic component. Similar processes take place
in the hierarchical scenario of structure formation, where massive objects
form in collisions and following mergings of less massive haloes. These processes
can be treated as collisions of the gas and dark matter flows. A collisionless
dark matter reaches the virial state apparently through the violent relaxation,
while in gas component shock waves form. The shock wave velocity
depends on mass of a forming protogalaxy: the velocity amplitude in a
perturbation of mass $M$ is close to the value
\begin{equation}
\label{vel}
v_c = \sqrt{3}\sigma,
\end{equation}
where $\sigma$ is the one-dimensional velocity dispersion \cite{blanchard}
\begin{equation}
\label{velvir}
  \sigma^2 = {GM\over 2R}.
\end{equation}
Thus, the parameters of the shock waves are determined by the total mass of matter involved
in motion, by the redshift at which the object forms, and so on. The efficiency
of HD formation is sensitive to these parameters. Therefore, one can expect
that the characteristcs of stellar population vary in galaxies of different mass.
As the maximum abundance of HD depends on the shock velocity, and
HD molecules cool the gas to much lower temperature than H$_2$ molecules do,
a typical mass of protostellar molecular clouds is expected to decrease
with increasing mass of a forming galaxy \cite{vir05}.

Shock waves in the epoch of galaxy formation can be connected
with explosions of the first supernovae. In these events much more powerful shock waves
form: typical velocities can be greater $\simgt 100$~km~s$^{-1}$,
the corresponding temperature behind the front is $\simgt 2.8 \times 10^5$~K. At
radiative stages when the gas temperature reaches $\simlt 10^4$~K,
the H$_2$ fraction becomes sufficiently high $\simgt 5\times 10^{-3}$ \cite{shapiro,ferrara98}
due to high ionization fraction at the preceding stages. As a consequence,
the gas temperature definitely falls to the lower values at which the cooling is essentially
determined by HD molecules. In these conditions fragmentation of an expanding shell
can occur \cite{vishniac83}.

Due to isobaric compression the gas density behind the shock increases
considerably compared to the initial value. For instance, for the velocity
$v_c \ge 7\alpha^{-1}$~km~s$^{-1}$
the initial temperature is $T \ge 5.8\times 10^3$~K (\ref{tshock}),
and when the gas cools down to $T_{\rm CMB}\simeq 2.7(1+z)$, its density increases
more than 200 times. At such conditions fragmentation and formation of stars
become possible behind the shock \cite{yamada}.
Gravitationally unstable fragments can give rise to protostars
or protostellar clusters. It follows that when cooling is determined by HD molecules,
formation of low mass protostellar clouds becomes possible \cite{uehara,nakamura}.
Indee, under these conditions the gas temperature falls down to $\sim 2.7(1+z)$~K
which at $z=20$ is 4 times smaller than can be provided only by H$_2$ molecules.
The Jeans mass $M_J \simeq 30T^{3/2}n^{-1/2} \msun$ behind the front is
$M_J\simeq 15T^{2}n_0^{-1/2}T_0^{-1/2}\msun$, where $n_0$, the gas density
in a flow (a cloud) before collision, $T_0$ is the temperature at the shock;
here density at the front is taken 4$n_0$ as for strong shock waves.
If we assume that gas density in flows (clouds) before collision is equal to the
virial value (see e.g. [24]), $n_0=18\pi^2n_b(1+z)^3$, then for velocities
$v_c \ge 7\alpha^{-1}$~km~s$^{-1}$ (or initial temperature
$T_0 \ge 5.8\times 10^3$~K) the Jeans mass is
\begin{eqnarray}
\label{mj}
 M_J\simlt 2.4\times 10^5\msun \left( {1+z\over T_0 \delta_c} \right)^{0.5}
   = 7.2\times10^3\msun \left({\alpha v \over 1~{\rm km~c^{-1}}}\right)^{-1}
                     \left({\delta_c \over 18\pi^2}\right)^{-0.5}
                     \left({1+z \over 20}\right)^{0.5}
\end{eqnarray}
where $\delta_c$ is the ratio of gas density before collision to the background baryonic
density. At the same time, when cooling is determined only by H$_2$ molecules the typical
gas temperature is of $\sim 200$~K, and the corresponding Jeans mass
$\sim 13.5[(1+z)/20]^{-2}$ times exceeds the value given by (\ref{mj})
\begin{eqnarray}
\label{mj200}
 M_J\simlt 1.3\times 10^9\msun \left( {1\over T_0 \delta_c (1+z)^3 } \right)^{0.5}
   = 10^5\msun \left({\alpha v \over 1~{\rm km~c^{-1}}}\right)^{-1}
                     \left({\delta_c \over 18\pi^2}\right)^{-0.5}
                     \left({1+z \over 20}\right)^{-1.5}
\end{eqnarray}

In other words, the question of how massive are the fragments formed
behind shock fronts depends on whether the cooling is determined by H$_2$ or
HD molecules. The density
of the fragments is $\ge 10 - 300$~cm$^{-3}$ depending on the redshift and the initial
temperature. Subsequent collapse is isothermal until the fragment
becomes opaque in H$_2$ and HD lines, which takes place at the density
$\sim 10^9-10^{10}$~cm~$^{-3}$.
At this stage, if the cooling is dominated by HD molecules, the Jeans mass is
$M_J\sim 30T_{\rm CMB}^{3/2}n^{-1/2}\msun\sim 10^{-3}(1+z)^{3/2}\msun$ \cite{vir05},
however when HD is underabundant the Jeans mass can be 2-3 orders
greater \cite{ferrara04}. Further evolution is determined by the accretion of gas
onto the central core \cite{omukaipalla,tan}. If the accretion rate is below the
Eddigton limit, the mass of a forming star is comparable to the initial mass of
a protostellar cloud, in the opposite case it can be much lower \cite{ferrara04,omukaipalla}.
Thus, one can expect that a typical
mass of the first stars born in protogalaxies of higher masses (corresponding to
higher collisional velocities) is shifted towards the lower end due
to cooling by HD molecules.

It is readily seen that since the overall thermal evolution of low density
flows differs from that of higher densities, the final value of the Jeans mass
and its dependence
on redshift will differ from the above value. Fig.7 shows the Jeans mass versus
the gas density in the flow. In the range of collisional velocities
$5.8\alpha^{-1}\leq  v_c < 8.6\alpha^{-1}$~km~s$^{-1}$
only flows with the initial density very close to the virial
can have Jeans mass $M_J$ smaller than $10^4\msun$,
which may correspond to the mass of a protostellar cloud. However for
$v \geq 8.6\alpha^{-1}$~km~s$^{-1}$ the Jeans mass becomes $\leq 10^4\msun$
for the initial density 4 times lower than the virial. As mentioned above,
for higher velocities
gas cools down to the CMB temperature, and the Jeans mass is $\leq 10^3\msun$,
what is seen also from (\ref{mj}) -- on Fig.7 flat parts of the lines for
the velocity $\geq 8.6\alpha^{-1}$~km~s$^{-1}$ reflects this circumstance.
Thus, the Jeans mass is considerably higher than (\ref{mj}) only for flows with a low
collisional velocity and a low density. High-velocity collisions $v_c\geq 8.6\alpha^{-1}$~km~s$^{-1}$ provide cooling down to
the temperature $T\simeq T_{\rm CMB}$ even for low density flows.

Let us consider collision of flows whose density is equal to the background value,
i.e. $\delta_c \simeq 1$, $\rho/\rho_{\rm vir} \simeq 6\times 10^{-3}$.
It is seen from Fig.7 that even for high-velocity collisions $v_c =
11.6\alpha^{-1}$~km~s$^{-1}$ gas cannot cool sufficiently in one Hubble
time, and the Jeans mass is quite high: $\sim 10^6-10^7~\msun$.
Moreover, the free-fall time for such low densities is greater than the
comoving Hubble time. Under these conditions baryonic objects cannot be formed.
However, further increase of the collisional velocity,
$v_c > 11.6\alpha^{-1}$~km~s$^{-1}$, makes the gas behind the shock able
to cool down to the temperature $\leq 1000$~K during one Hubble time.
For instance, for low density flows collided with the
velocity $v_c \simeq 19.2\alpha^{-1}$~km~s$^{-1}$ the final
temperature behind the shock is $\sim 200$~K, and can reach
lower values for higher velocities. Under these conditions HD molecules
will dominate in gas colling, and the Jeans mass becomes as small as
$\sim ~7\times 10^4\msun$.

%----------------------- Section 4 -------------------------------
\section{Conclusions}

\noindent

The influence of HD molecules on the thermochemical evolution of
the primordial gas behind shock waves possibly formed during
the epoch of galaxy formation has been studied.

\begin{enumerate}
\item We showed that deuterium converts efficiently to HD molecules
and the contribution of HD to coolig becomes dominant for the shock waves with velocities
$\simgt 4.6\alpha^{-1}$~km~s$^{-1}$ ($\alpha \simeq 0.5$).
Behind such shock waves the conditions are favourable for
fragmentation and, as a consequence, for formation of protostellar clusters.

\item For shock velocities $\simgt 7\alpha^{-1}$~km~s$^{-1}$ gas is able to
cool down to the CMB temperature. Under these conditions
Jeans mass depends only on the redshift and the initial density:
$M_J\simlt 2.4\times 10^5\msun (1+z)^{0.5} (T_0 \delta_c)^{-0.5}$,
for virial haloes ($\delta_c = 18\pi^2$) at $z =20$ this corresponds to
$M_J\simlt 10^3\msun$.

\item At $z\simgt 45$ the CMB temperature is close to the critical value
$T_{\rm cr}$, at which
the contribution from HD molecules to the total cooling is comparable to
that from H$_2$. At these conditions HD molecules begin to heat gas, and
at higher redshifts become unimportant in thermal history of baryons.

\item For densities of colliding flows smaller than the virial value the efficiency
of HD molecule formation decreases. In particular, at $z=20$ gas temperature
behind the shock with $v\sim 5.8\alpha^{-1}$~km~s$^{-1}$ drops substantially
only for a density close to the virial value. However, for the shock
velocities $\sim 8.6\alpha^{-1}$~km~s$^{-1}$ HD molecules are important
in cooling for densities of 2-3 times lower than the virial value.
For $v\sim 11.6\alpha^{-1}$~km~s$^{-1}$ HD cooling is effective
for densities close to the background value, and almost all deuterium converts
to HD for collisions of less dense than if the flows were virial.
For the gas density equal to the
background value and for the velocity $v\simgt 19.2\alpha^{-1}$~km~s$^{-1}$
temperature drops to $\leq 200$~K and HD molecules begin to dominate radiative
cooling.

\end{enumerate}

%----------------------- Section L -------------------------------

\vskip 1.5cm
{\it Note added in manuscript 2005 July 26.-} After acceptance of this
paper we have been informed about the paper by A. Lipovka, 
R. N\'u\~nez-L\'opez, and V. Avila-Reese (MNRAS, 2005,
in press, astro-ph/0503682), where new calculations of the
HD cooling function are reported. In the temperature range of 
interest ($T<10^3$ K) this function coincides with that given 
in [18], while at higher temperatures, where Lipovka et al 
predict an order of magnitude enhanced HD cooling rate, the 
abundance of HD is too low to contribute. 

\newpage

%------------------------ figures --------------------------------

%%%%%%%%%%%%%%%%%%%%%%%%%%%%%%%%%%%%%%%%%%%%%%%%%%%%%%
\begin{figure}
\epsfxsize=12cm
\epsfbox{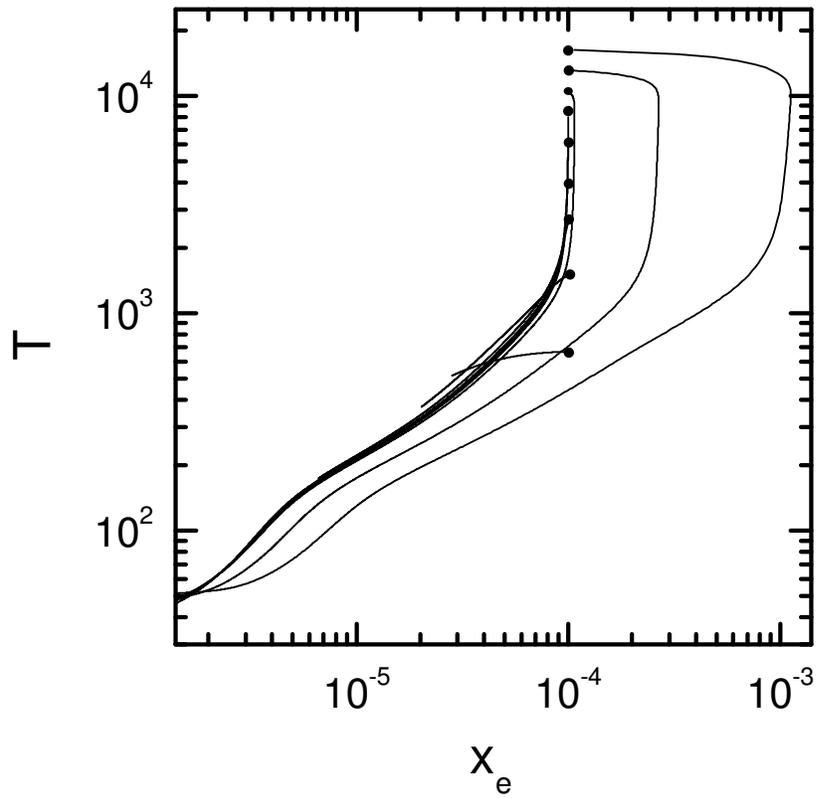}
\caption{
The evolutionary paths connecting fractional ionization and
gas temperature behind shocks in collisions with velocities
2.3, 3.5, 4.6, 5.8, 7, 8, 9.2, 10.4, 11.6~$\alpha^{-1}$~km~s$^{-1}$ -- from top
to bottom; initial points of the evolution are marked by filled circles.
}
\label{ele}
\end{figure}
%%%%%%%%%%%%%%%%%%%%%%%%%%%%%%%%%%%%%%%%%%%%%%%%%%%%%%

%%%%%%%%%%%%%%%%%%%%%%%%%%%%%%%%%%%%%%%%%%%%%%%%%%%%%%
\begin{figure}
\epsfxsize=12cm
\epsfbox{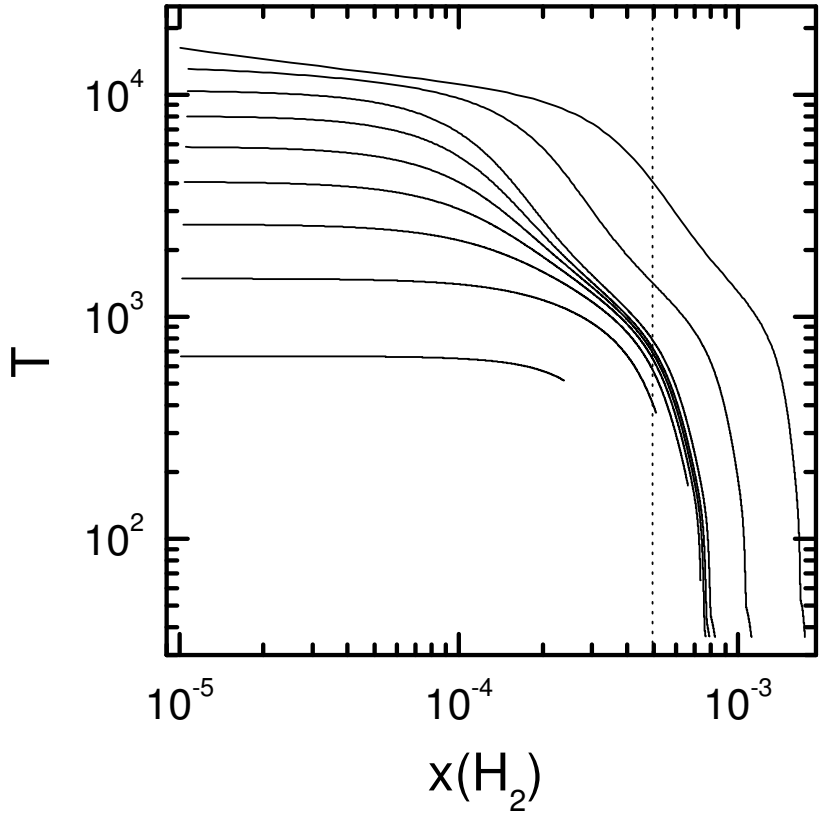}
\caption{
The evolutionary paths connecting variations of temperature
and H$_2$ concentration for the same velocities as in Fig. 1
(left to right). The vertial line depicts the critical value
$5\times 10^{-4}$ needed for baryons to cool in one comoving
Hubble time \cite{t97}.
}
\label{h2}
\end{figure}
%%%%%%%%%%%%%%%%%%%%%%%%%%%%%%%%%%%%%%%%%%%%%%%%%%%%%%

%%%%%%%%%%%%%%%%%%%%%%%%%%%%%%%%%%%%%%%%%%%%%%%%%%%%%%
\begin{figure}
\epsfxsize=12cm
\epsfbox{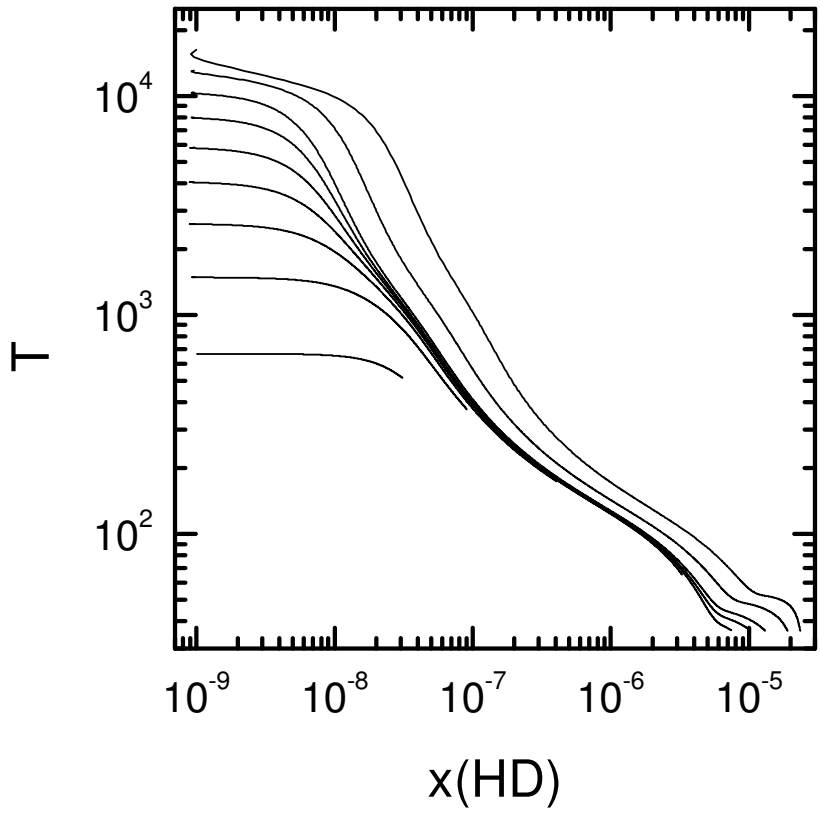}
\caption{
The evolutionary paths for the abundance of HD molecules (see Fig. 1).
}
\label{hd}
\end{figure}
%%%%%%%%%%%%%%%%%%%%%%%%%%%%%%%%%%%%%%%%%%%%%%%%%%%%%%
%%%%%%%%%%%%%%%%%%%%%%%%%%%%%%%%%%%%%%%%%%%%%%%%%%%%%%
\begin{figure}
\epsfxsize=12cm
\epsfbox{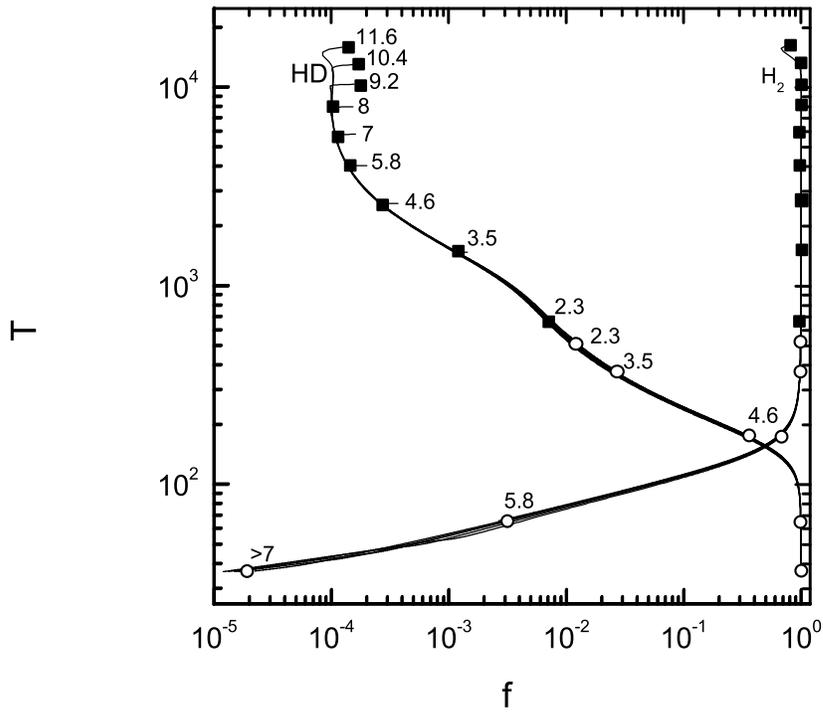}
\caption{
The relative contribution of radiative losses in the lines of
H$_2$ and HD into the total cooling (see Fig.1). Squares mark the initial points,
circles -- the final points of the evoluton for the velocities
2.3, 3.5, 4.6, 5.8, 7, 8, 9.2, 10.4, 11.6 $\alpha^{-1}$~km~s$^{-1}$.
}
\label{fh2temp}
\end{figure}
%%%%%%%%%%%%%%%%%%%%%%%%%%%%%%%%%%%%%%%%%%%%%%%%%%%%%%
%%%%%%%%%%%%%%%%%%%%%%%%%%%%%%%%%%%%%%%%%%%%%%%%%%%%%%
\begin{figure}
\epsfxsize=15cm
\epsfbox{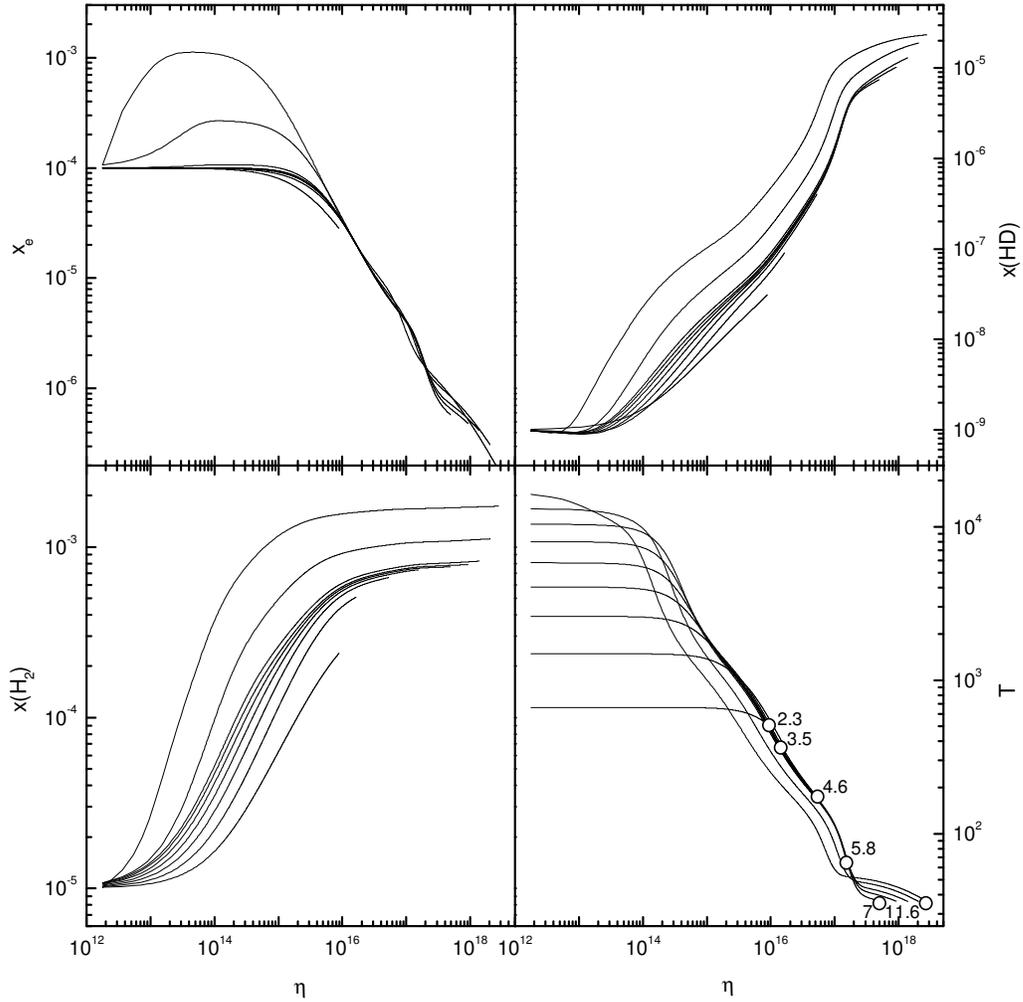}
\caption{
Ionization, H$_2$, HD fractions and temperature behind the shock front
versus $\eta$ for the velocities
2.3, 3.5, 4.6, 5.8, 7, 8, 9.2, 10.4, 11.6~$\alpha^{-1}$~km~s$^{-1}$.
Circles on temperature lines mark the final points of the evolution.
}
\label{evol}
\end{figure}
%%%%%%%%%%%%%%%%%%%%%%%%%%%%%%%%%%%%%%%%%%%%%%%%%%%%%%
%%%%%%%%%%%%%%%%%%%%%%%%%%%%%%%%%%%%%%%%%%%%%%%%%%%%%%
\begin{figure}
\epsfxsize=12cm
\epsfbox{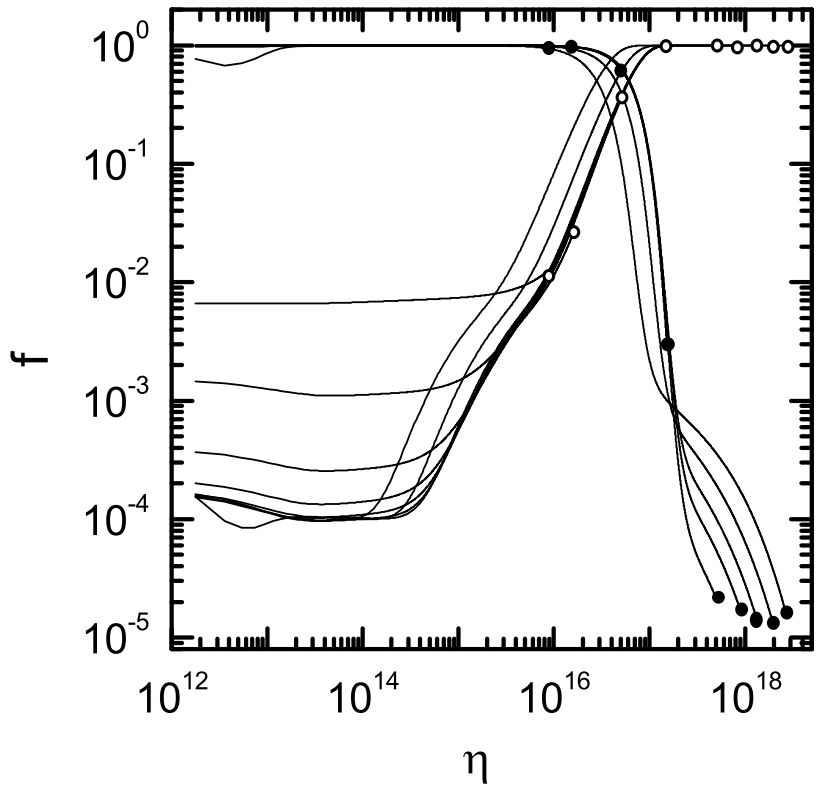}
\caption{
The relative contribution of radiative losses in the lines of
H$_2$ and HD into the total cooling (see Fig.5). Filled (H$_2$) and open (HD) circles
mark final points of the evoluton for velocities
2.3, 3.5, 4.6, 5.8, 7, 8, 9.2, 10.4, 11.6 $\alpha^{-1}$~km~s$^{-1}$ (left to right).
}
\label{radcontibution}
\end{figure}
%%%%%%%%%%%%%%%%%%%%%%%%%%%%%%%%%%%%%%%%%%%%%%%%%%%%%%
%%%%%%%%%%%%%%%%%%%%%%%%%%%%%%%%%%%%%%%%%%%%%%%%%%%%%%
\begin{figure}
\epsfxsize=9.5cm
\epsfbox{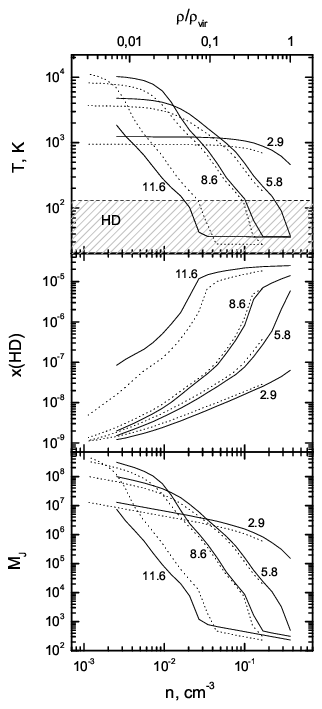}
\caption{
Temperature, HD fraction and Jeans mass reached behind the shock fronts
with the velocities 2.9, 5.8, 8.6, 11.6 $\alpha^{-1}$~km~s$^{-1}$
for two redshifts: $z = 20$ -- solid, $z = 15$ -- dashed lines.
Shaded area on the top panel shows the region where the HD cooling prevails.
The density in units of the virial value for $z = 20$ is shown at the upper axis.
}
\label{evoldenfig}
\end{figure}
%%%%%%%%%%%%%%%%%%%%%%%%%%%%%%%%%%%%%%%%%%%%%%%%%%%%%%


\begin{thebibliography}{99}

\bibitem{lepp84} S. Lepp, J.M. Shull, Astrophys. J. {\bf 280}, 465 (1984)

\bibitem{puy93} D. Puy, G. Alecian, J. Le Bourlot, et al., Astron. and Astrophys. {\bf 267}, 337 (1993)

\bibitem{palla95} F. Palla, D. Galli, J. Silk, Astrophys. J. {\bf 451}, 44 (1995)

\bibitem{gp} D. Galli, and F. Palla, Astron. and Astropys. {\bf 335}, 403 (1998)

\bibitem{sld98} P.C. Stancil, S. Lepp, A. Dalgarno, Astrophys. J. {\bf 509}, 1 (1998)

\bibitem{bgalli97} E. Bougleux, D. Galli, Monthly Notes Roy. Soc., {\bf 288}, 638 (1997)

\bibitem{puy97} D. Puy, and M. Signore, NewA {\bf 2}, 299 (1997)

\bibitem{puy} D. Puy, and M. Signore, NewA {\bf 3}, 247 (1998)

\bibitem{palla00} F. Palla, Proceedings of Star Formation 1999, ed. T. Nakamoto,
 Nobeyama Radio Observatory,  p.6 (1999)

\bibitem{coll} A. A. Suchkov, Yu. A. Shchekinov, M. A. Edelman, Astrophysics, {\bf 18}, 360 (1983)

\bibitem{maclow86} M.-M. Mac Low, J.M. Shull, Astrophys. J. {\bf 302}, 585 (1986)

\bibitem{shapiro} P.R. Shapiro, H. Kang, Astrophys. J. {\bf 318}, 32 (1987)

\bibitem{kang} H. Kang, P.R. Shapiro, Astrophys. J. {\bf 386}, 432 (1992)

\bibitem{solomon} P. M. Solomon, N. J. Woolf, Astrophys. J. {\bf 180}, 89 (1973)

\bibitem{varsh} D. A. Varshalovich, V. K. Khersonskii, SvAL, {\bf 2}, 227 (1976)

\bibitem{wmap} D. N. Spergel, L. Verde, H.V. Peiris et al, Astophys. J. Suppl. {\bf 148}, 175 (2003)

\bibitem{hm} D. Hollenbach and C.F. McKee, Astrophys. J. Suppl. {\bf 41}, 555 (1979)

\bibitem{flower} D. Flower, Monthly Notes Roy. Soc., {\bf 318}, 875 (2000)

\bibitem{cen92} R. Cen, Astron. J. Suppl. {\bf 78}, 341 (1992)

\bibitem{gp00} D. Galli, and F. Palla, Planetary and Space Sci. {\bf 12-13}, 1197 (2002)

\bibitem{smith} J. Smith, Astrophys. J. {\bf 238}, 842 (1980)

\bibitem{annorman} P. Anninos, M. Norman, Astrophys. J. {\bf 460}, 556 (1996)

\bibitem{zeldnov} Ya. B. Zeldovich, I.D. Novikov, Relativistic astrophysics. v.2 - The structure and evolution of the universe, University of Chicago Press (1983)

\bibitem{t97} M. Tegmark, J. Silk, M.J. Rees, et al., Astrophys. J. {\bf 474}, 1 (1997)

\bibitem{ohhaiman} S.P. Oh, Z. Haiman, Astrophys. J. {\bf 569}, 558 (2002)

\bibitem{lin} C. Lin, L. Mestel, F. Shu,  Astrophys. J. {\bf 142}, 1431 (1965)

\bibitem{zeld} Ya. B. Zeldovich, Astron. \& Astrophys. 5, 84 (1970)

\bibitem{blanchard} A. Blanchard, D. Valls-Gabaud, G.A. Mamon, Astron. and Astrophys., {\bf 264}, 365 (1992)

\bibitem{vir05} E. O. Vasiliev, Yu. A. Shchekinov, Astr. Rept. 49, 587 (2005)

\bibitem{ferrara98} A. Ferrara, Astophys. J., {\bf 499}, L17 (1998)

\bibitem{vishniac83} E. T. Vishniac, Astrophys. J., {\bf 274}, 152 (1983)

\bibitem{yamada} M. Yamada, R. Nishi, Astrophys. J. {\bf 505}, 148 (1998)

\bibitem{uehara} H. Uehara, S. Inutsuka, Astrophys. J. Lett. {\bf 531}, 91 (2000)

\bibitem{nakamura} F. Nakamura, M. Umemura, Astrophys. J. {\bf 569}, 549 (2002)

\bibitem{ferrara04} B. Ciardi, A. Ferrara, Space Sci Rev. {\bf 116}, 625

\bibitem{omukaipalla} K. Omukai, F. Palla, Astrophys. J. {\bf 589}, 677 (2003)

\bibitem{tan} J. Tan, C.F. McKee, Astrophys. J. {\bf 603}, 383 (2004)


%\bibitem{}


\end{thebibliography}
\end{document}